\documentclass[12pt,preprint]{aastex}
\usepackage{amsmath}
\usepackage{natbib}

\begin{document}

\bibliographystyle{apj}

\title{{\it MOST}\footnote{Based on data from the {\it MOST} satellite, 
a Canadian Space Agency mission, jointly operated by Dynacon
Inc., the University of Toronto Institute for Aerospace Studies and the
University of British Columbia, with the assistance of the University of
Vienna.}  Spacebased Photometry of the Transiting Exoplanet System HD 189733: 
Precise Timing Measurements for Transits Across an Active Star}

\author{Eliza Miller-Ricci}

\affil{Harvard-Smithsonian Center for Astrophysics, 60 Garden St. Cambridge,
        MA 02138}

\email{emillerricci@cfa.harvard.edu}

\author{Jason F. Rowe}

\affil{University of British Columbia, 6224 Agricultural Road, Vancouver, BC
        V6T 1Z1, Canada}

\author{Dimitar Sasselov}

\affil{Harvard-Smithsonian Center for Astrophysics, 60 Garden St. Cambridge,
        MA 02138}

\author{Jaymie M. Matthews}

\affil{University of British Columbia, 6224 Agricultural Road, Vancouver, BC
        V6T 1Z1, Canada}

\author{Rainer Kuschnig}

\affil{Department of Physics and Astronomy, University of British Columbia,
        6224 Agricultural Road, Vancouver, BC V6T 1Z1, Canada}

\author{Bryce Croll}

\affil{Department of Astronomy and Astrophysics, University of Toronto, 50
        Saint George St., Toronto, ON M5S 3H4, Canada}

\author{David B. Guenther}

\affil{Department of Astronomy and Physics, St. Mary's University, Halifax, NS
        B3H 3C3, Canada}

\author{Anthony F.J Moffat}

\affil{D\'{e}partement de Physique, Universit\'{e} de Montr\'{e}al, C.P. 6128,
       Succ. Centre-Ville, Montr\'{e}al, QC H3C 3J7, Canada}

\author{Slavek M. Rucinski}

\affil{David Dunlap Observatory, University of Toronto, P.O. Box 360,
       Richmond Hill, ON L4C 4Y6, Canada}

\author{Gordon A.H Walker}

\affil{Department of Physics and Astronomy, University of British Columbia,
        6224 Agricultural Road, Vancouver, BC V6T 1Z1, Canada}

\author{Werner W. Weiss}

\affil{Institut f\"{u}r Astronomie, Universit\"{a}t Wien,
       T\"{u}rkenschanzstrasse 17, A-1180 Wien, Austria}

\begin{abstract}

We have measured transit times for HD 189733b passing in front of its bright 
(V = 7.67) chromospherically active and spotted parent star.  
Nearly continuous broadband optical photometry of this system was obtained  
with the {\it MOST} (Microvariability \& Oscillations of STars) space 
telescope during 21 days in August 2006, monitoring 10 consecutive transits.  
We have used these data to search for deviations from a constant 
orbital period which can indicate the presence of additional planets in the 
system that are as yet undetected by Doppler searches.  There are no transit 
timing variations above the level of ${\pm}45$ s, ruling out super-Earths 
(of masses $1 - 4 M_{\earth}$) in the 1:2 and 2:3 inner resonances and
planets of 20 $M_{\earth}$ in the 2:1 outer resonance of the 
known planet.  We also discuss complications in measuring transit times for a 
planet that transits an active star with large star spots, and how the 
transits can help constrain and test spot models. This has implications for 
the large number of such systems expected to be discovered by the CoRoT and 
Kepler missions.    

\end{abstract}


\section{Introduction}

While groundbased radial velocity (RV) and photometric transit surveys have 
unearthed
more than 200 extrasolar planets in just over the last decade\footnote{The
Extrasolar Planets Encyclopedia: http://www.exoplanet.eu.}, the ability to 
detect planets similar to the Earth, in size and mass, has so far remained out
of reach of both of these methods.   However, the potential to discover 
planets comparable to the mass of the Earth currently exists through 
measurements of 
transit timing variations (TTVs) in known transiting 
planetary systems \citep{ago05, hol05, hey06}.  Additionally, nearly 
continous photometry from space currently 
offers the precision and time coverage to 
search for Earth-size transiting planets (e.g., Barge et al.~2005, 
Croll et al.~2007).  

A summary of the TTV method is as follows.  
The motion of a transiting planet whose orbit is perturbed by another 
planet in the system will not have a constant period.  
These changes in period occur on the level of seconds to hours
\citep{ago05, hol05} depending on the mass and orbital parameters of the
pertubing planet.  By measuring the times for repeated transits
of a known planet, one can at least constrain, if not unambiguously determine,
the mass and semi-major axis of the perturbing body.  
Such TTV analyses can be performed on any exoplanet for which high quality 
photometry is available for a number of transits.  
For the TrES-1 system an analysis of transit timings has been
carried out by \citet{ste05} with the result that any companion planet
in an orbit nearby to the known close-in giant planet must have a mass 
comparable to or less than that of Earth.  For the HD 209458 system, 
\citet{ago07} and \citet{mil07}
have placed limits on the existence of additional low-mass planets down
to less than an Earth mass in certain resonant orbits, using data from 
the Hubble Space Telescope (HST) and the Microvariability
and Oscillations of STars ({\it MOST}) \citep{wal03, mat04} 
satellite, respectively.
Here we apply a similar TTV analysis to the HD 189733 system to search for
low-mass companion planets in orbits neighboring the known 
transiting planet.

HD 189733 is currently the brightest star (V = 7.67) known to harbor a 
transiting exoplanet \citep{bou05}.  This fact along with its position on the 
sky and the short (2.2 d) period of its transiting planet makes it ideally 
suited for observations by the {\it MOST} satellite. 
{\it MOST} observed HD 189733 for 21 days in August 2006, 
monitoring 10 consecutive planetary transits.  
By determining the timings of these transits we are able to place limits on 
the presence of additional planets in this system down to a level of several 
Earth masses in certain orbits.    

Another characteristic of HD 189733 that makes it unique among the 
known transiting systems is that the star has surface spots which modulate 
its optical light at a level of about 3\% during the stellar rotation cycle  
(Winn et al.\ 2006a,  Matthews et al.\ in preparation). This level 
of variability is consistent with the fact that the star is known to be 
relatively active, with a chromospheric activity index $S = 0.525$ 
\citep{wri04}.  The presence of star spots on HD 189733 must be taken 
into account when fitting the light curves of the exoplanetary transits, 
since the shapes of individual transits can be affected (1) if they 
coincide with the planet's passage in front of a spot, or (2) if the 
underlying variability of the star is rapid enough to affect the fit to the 
transit.

The CoRoT \citep{bag03, bar05} and Kepler \citep{bor04, bas05} 
satellite missions (launched December 27, 2006 and scheduled for launch in 
February 2009, respectively) are both expected to detect a large number of new
transiting planets to add to an already growing list.  One exciting prospect of
these new systems is that each of them should be suited to the type of
TTV analysis that we present in this paper, since high-quality
observations of multiple consecutive transits will be available.  
Many of these newly discovered exoplanets may transit in front of 
chromospherically active stars like HD 189733, since such stars are 
commonplace, and neither the CoRoT nor Kepler team has done {\em a priori} 
selection of quiet stars for their transit searches.  In the following
TTV analysis, we incorporate the intrinsic variability of HD 189733, 
in an attempt to quantify the effects that this may have on the measurement 
of transit times.  This provides valuable experience for future 
observations of transits of active spotted stars by CoRoT, Kepler, 
and groundbased surveys.

\section{{\it MOST} Photometry}

HD 189733 was observed by {\it MOST} nearly continuously for 21 days from
July 31 to August 21, 2006.  The photometry was collected in MOST's Direct 
Imaging mode operating with a single CCD, where a defocused stellar image was 
recorded on a subraster 
of the MOST Science CCD.  The exposure time for individual frames was 1.5 s
(read time is negligible), and 14 
consecutive images were "stacked" onboard the satellite to achieve high S/N 
with a time sampling interval of 21 s.  The original plan for the duration of 
the HD 189733 run was 14 days, and the first 
14 days of observation have a duty cycle of 94\%.  When examination of the 
light curve showed the obvious spot modulation, the run was extended for 7 
more days, by sharing each MOST satellite orbit ($P = 101.4$ min) with the 
next scheduled Primary Science Target. Therefore, the last week of data have a 
reduced duty cycle of 46\%, and the observations were restricted to MOST 
orbital phases of highest scattered Earthshine (with a resultant increase in 
photometric scatter).  Still, the time sampling and photometric 
precision for the final week of observations remain excellent and are 
sufficient for analysis of the transit light curve.

The data were reduced in the same way as described by \citet{row06} for 
the transiting system HD 209458.  The stellar fluxes 
were extracted by aperture photometry (aperture radius = 4 pixels; stellar 
image FWHM $\sim$ 2.5 pixels), which gives better results than PSF 
(Point-Spread-Function) fitting.  The raw instrumental light curve was 
decorrelated against the sky background as described by \citet{row06} 
and also against the location of the PSF centroid on the CCD. 

\begin{figure}
\plotone{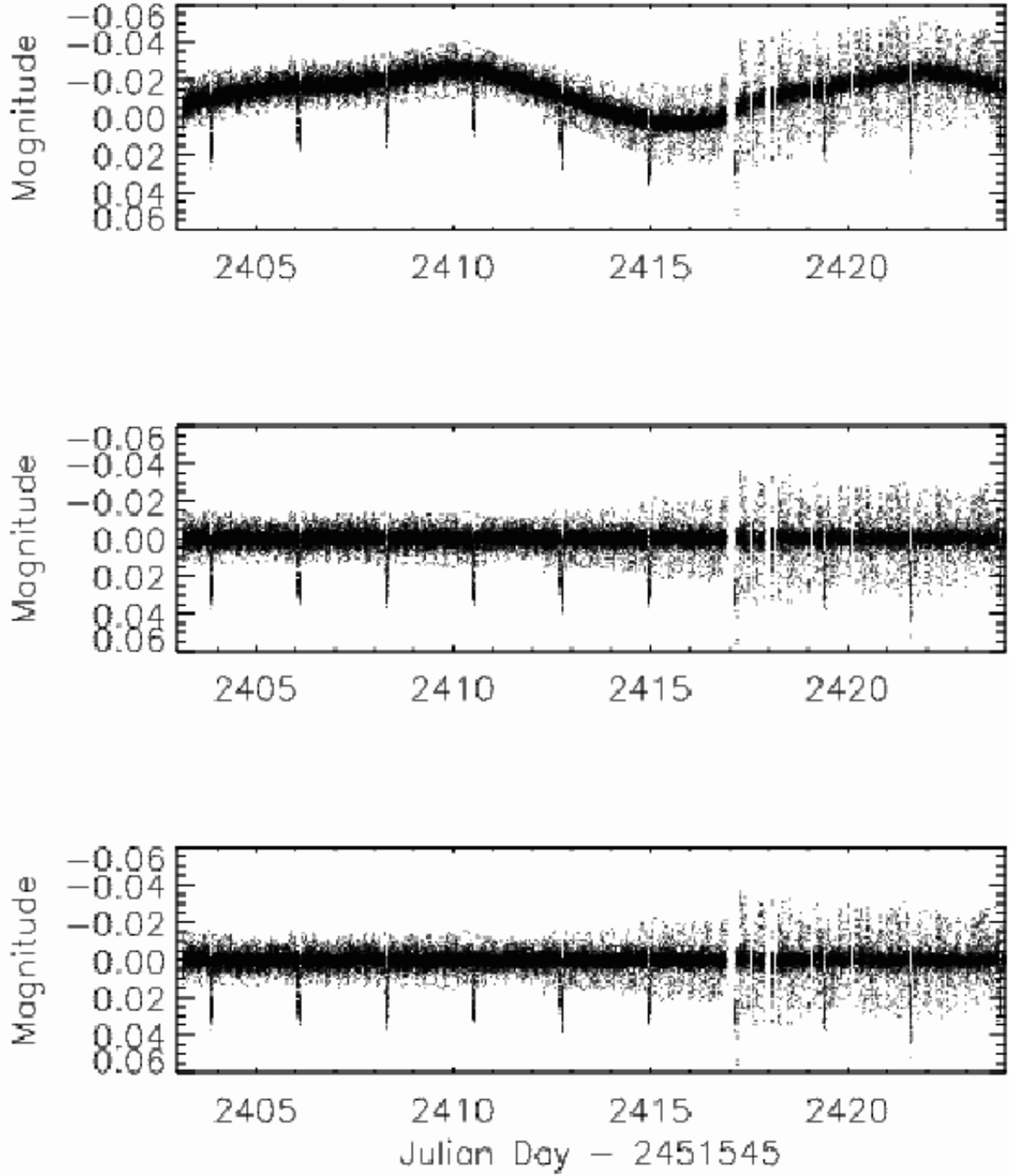}
\caption{Top: {\it MOST} light curve of HD 189733.  Middle: light curve 
        normalized by smoothing the OOT data and subtracting this smoothed
	curve.  Bottom: light curve normalized by applying a 
	filtered Fourier Transform to remove all power except at the orbital
	period of the planet and its harmonics.
        \label{lc}}
\end{figure}

The reduced light curve of HD 189733 is shown in the top panel of 
Figure~\ref{lc}.  It is immediately apparent that the star itself was 
variable during the {\it MOST} observations at a level of approximately 30 
mmag.  The shape and timescale of these variations are consistent with 
rotational modulation due to spots on the surface of HD 189733.  

\section{Fitting the Transit Light Curve}

To fit and time the observed transits properly, the underlying variability of 
the host star must be subtracted from the light curve.  We have adopted two 
independent approaches: (1) We smooth the out-of-transit (OOT) light curve 
using as a kernel a weighted average of the data covering one {\it MOST} 
satellite 
orbital period (about 101 min) centered on each data point. For the in-transit 
data, we linearly interpolate the smoothed data through the time of transit.  
Finally, we subtract the resulting smoothed light curve from the data. (2) 
We apply a filtered  Discrete Fourier Transform to the data, in which we 
remove all power except at the orbital period of the exoplanet HD 189733b and 
its harmonics.  We use the orbital period determined by \citet{win06b}.

The filtered light curves resulting from these two methods are shown in the 
middle and bottom panels of Figure~\ref{lc}, respectively.  In the subsequent 
TTV analysis, we use both of the independently normalized light curves and 
find consistent results from each.  

\subsection{Transit Model}\label{tr_model}

In computing the transit times, we compare the {\it MOST} data against a model
transit light curve, constructed using the formalism set forth in
\citet{man02}.  To 
determine the system parameters describing this model $-$ stellar 
mass and radius ($M_{*}$ and $R_{*}$), planetary
radius ($R_{p}$), orbital inclination ($i$), orbital period ($P$), and stellar
limb-darkening coefficients ($c_{1}-c_{4}$) $-$ we procede as 
follows.  

Since $M_{*}$ cannot be determined from photometry alone, we adopt 
the value from \citet{bou05} of $M_{*} = 0.82 \pm 0.03 M_{\sun}$, which in turn
was determined by fitting the star's observed spectral parameters against 
stellar evolution models.  For $R_{*}$, $R_{p}$, 
and $i$, we take advantage of the high-quality photometry provided by 
{\it MOST} to do an independent fit to these parameters.  
To determine the transit parameters for the HD 189733 system, we use a
maximum likelihood analysis to fit the data with the analytic model of 
\citet{man02} and also add reflected light from the planetary companion 
(see Matthews et al.\ (in preparation) for details).  For priors on the 
stellar and planetary mass 
and radius as well as the orbital period and inclination of the planet, we
adopt the values given in \citet{win06b}.  
We obtain stellar and planetary radii of 
$0.749 \pm 0.009$ R$_{\sun}$ and $1.192 \pm 0.019$ R$_{Jup}$ (equatorial 
radius), respectively, and we find the orbital inclination angle to be
$85.70^{\circ} \pm 0.11^{\circ}$.

For the above analysis, we fixed the orbital period of HD 189733b at 
2.2185733 $\pm$ 0.0000019 days
\citep{win06b}.  This value was determined by compiling
all available transit times for fully observed transits over a period of more
than a  year, making it more precise than what could be obtained from a fit 
to the {\it MOST} light curve alone.  In Section~\ref{analysis} we present our
own calculation of $P$ by adding our transit times to the list of those
previously available, and we find that the value we obtain is entirely 
consistent with the one we have chosen to use for our model transit 
light curve.

\begin{figure}
\plotone{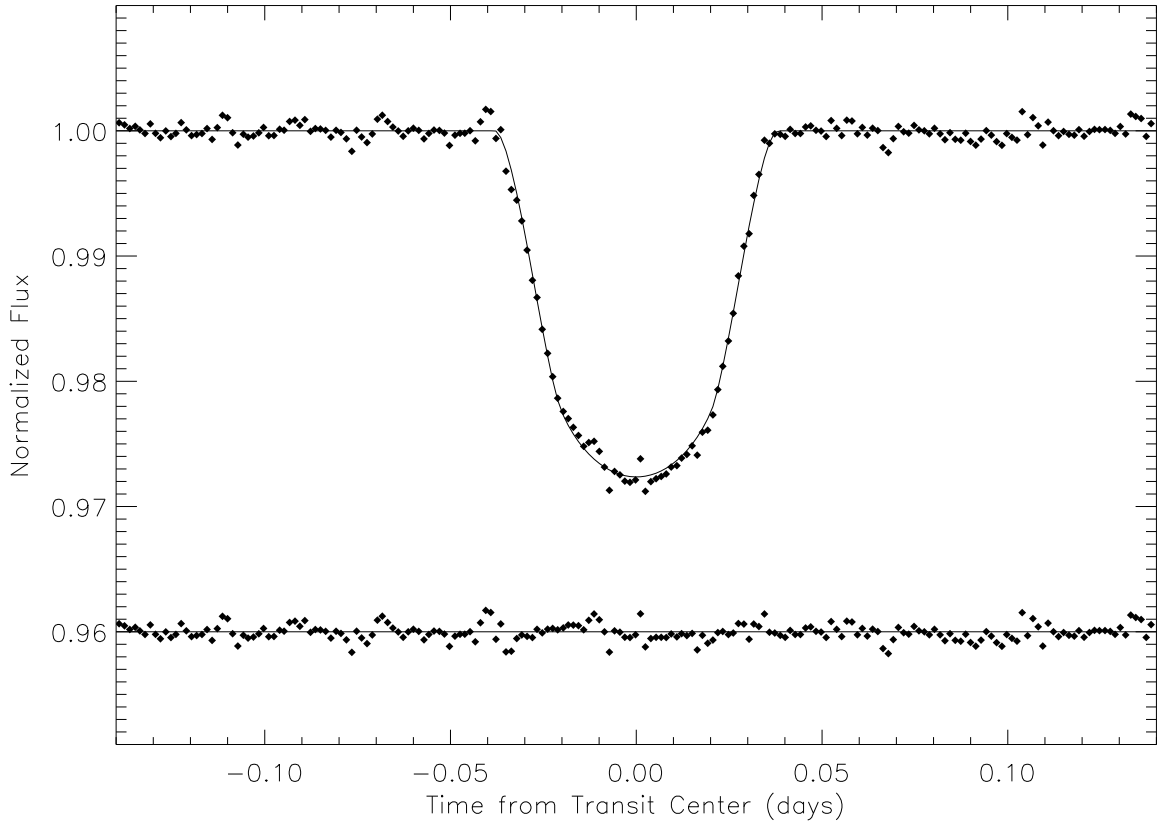}
\caption{Above: The phase diagram of the {\it MOST} photometry of HD 189733, 
        and the transit model, folded at the orbital period of the planet. The 
	data have been averaged in 2-min bins. Below: Residuals from the 
	model. \label{phase}}
\end{figure}

Since the {\it MOST} telescope has a custom broadband filter covering the
range $350 -  700$ nm, we cannot use standard tables of limb-darkening
coefficients (such as those by \citet{cla00}), which are only valid for
standard filter systems.  In addition, the quality of the {\it MOST} photometry
is not high enough to observationally constrain limb-darkening parameters, 
especially given the wide bandpass of the instrument.  
We instead generate non-linear limb darkening
parameters for HD 189733 from a synthetic spectrum, calculated
from model atmospheres using the ATLAS 9 and ATLAS 12 codes by 
\citet{kur95} and rewritten in FORTRAN 90 (J. Lester, private communication).
We employ a model with T$_{eff}$ = 5000 K, $\log g = 4.5$, microturbulence
of 1 km/s, and solar metallicity, consistent with the stellar parameters for 
HD 189733 given by \citet{bou05} and \citet{mel06}.  The 
emergent spectrum is calculated at each of 17 values of 
$\mu = \cos \theta$ from the center of the star out to the limb, where 
$\theta$ is the angle relative to the normal.  We then
determine non-linear limb-darkening coefficients ($c_{n}$) as a function of 
wavelength according to the law:
\begin{equation}
I(r) = \sum_{n=1}^{4}c_{n}(1-\mu^{n/2}) \ .
\end{equation}
Finally, we multiply by the total throughput function for {\it MOST}, which 
includes the effects of the filter, optics, CCD, and its electronics.  
The resultant
limb-darkening coefficients for the {\it MOST} bandpass, along with the other 
system parameters for the 
model light curve, are listed in Table~\ref{tbl1}.  In Figure~\ref{phase}, 
the transit model is 
plotted over the binned (normalized) data, which have been folded at the
orbital period of HD 189733b. 

\subsection{Transit Times}

Using the transit model described above, we find the best-fit center-of-transit
times for 9 of the 10 transits observed by {\it MOST}.  Transit 7
is omitted due to the
fact that {\it MOST} was switching observing modes at this time to the 
dual-star mode.  For the 9 transits that have good time coverage, we 
compute the model light curve for times that
correspond with each of the {\it MOST} data points.  We next determine the
center time for each transit at which the $\chi^2$ value for the
fit to the data is minimized.  

The 1-$\sigma$ 
error bars are calculated using a bootstrapping Monte Carlo simulation similar
to the one
described in \citet{ago07}, according to the following sequence.  For each 
transit we
shift the residuals from the best-fitted transit model  (and their associated
errors) by a random number of points.  We then add the new residuals
onto the transit model and recalculate the center-of-transit time using the
same procedure described above, thus maintaining the original point-to-point
correlations.  We perform this procedure 200 times for each transit, and the
scatter in the obtained transit times sets our uncertainties.
This allows us to account for the effects of correlated (red) 
noise in the {\it MOST} data set, which is due to the repeating pattern of 
stray light (satellite orbital modulation of scattered 
Earthshine) as well as to other systematics that remain in the light curve 
including any residual intrinsic stellar variations. 

\begin{figure}
\plotone{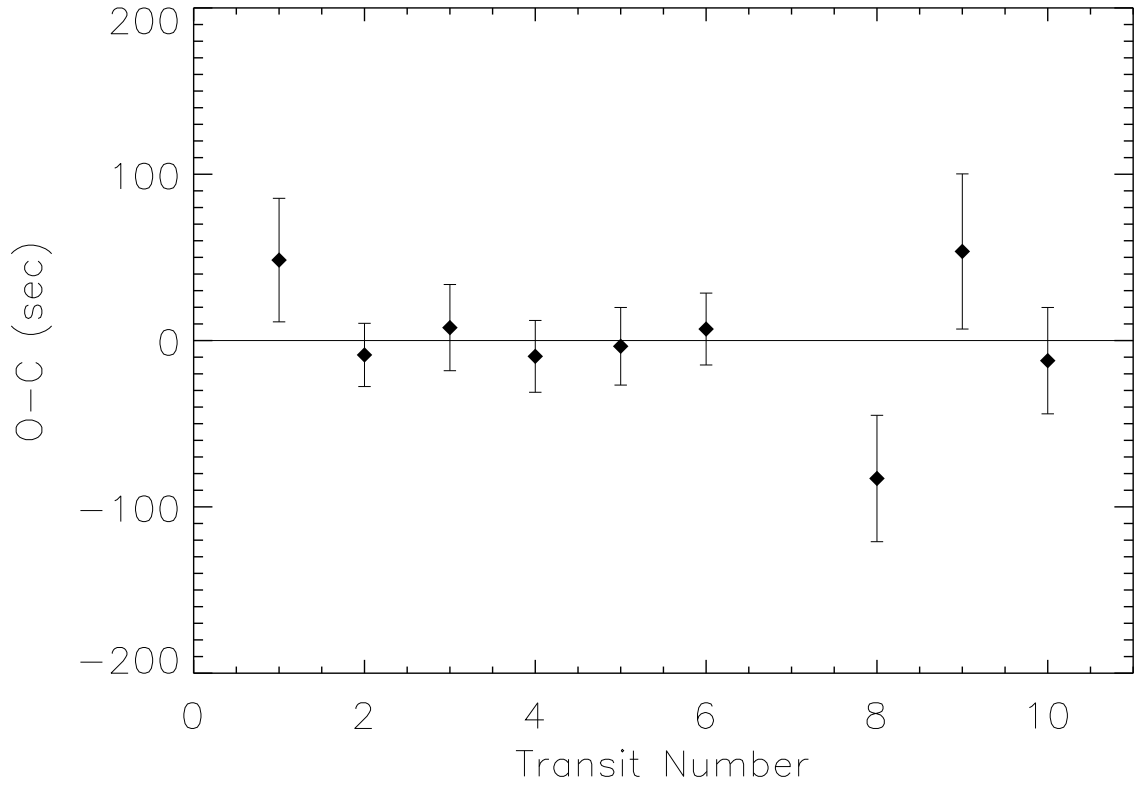}
\caption{Deviation from predicted time of transit vs. transit number for
        transits of HD 189733b observed by {\it MOST} in 2006.  The first
	6 transits were fully sampled, while transits 8, 9, and 10 were
	only partially observed.
        \label{O-C_2006}}
\end{figure}

\begin{figure}
\plotone{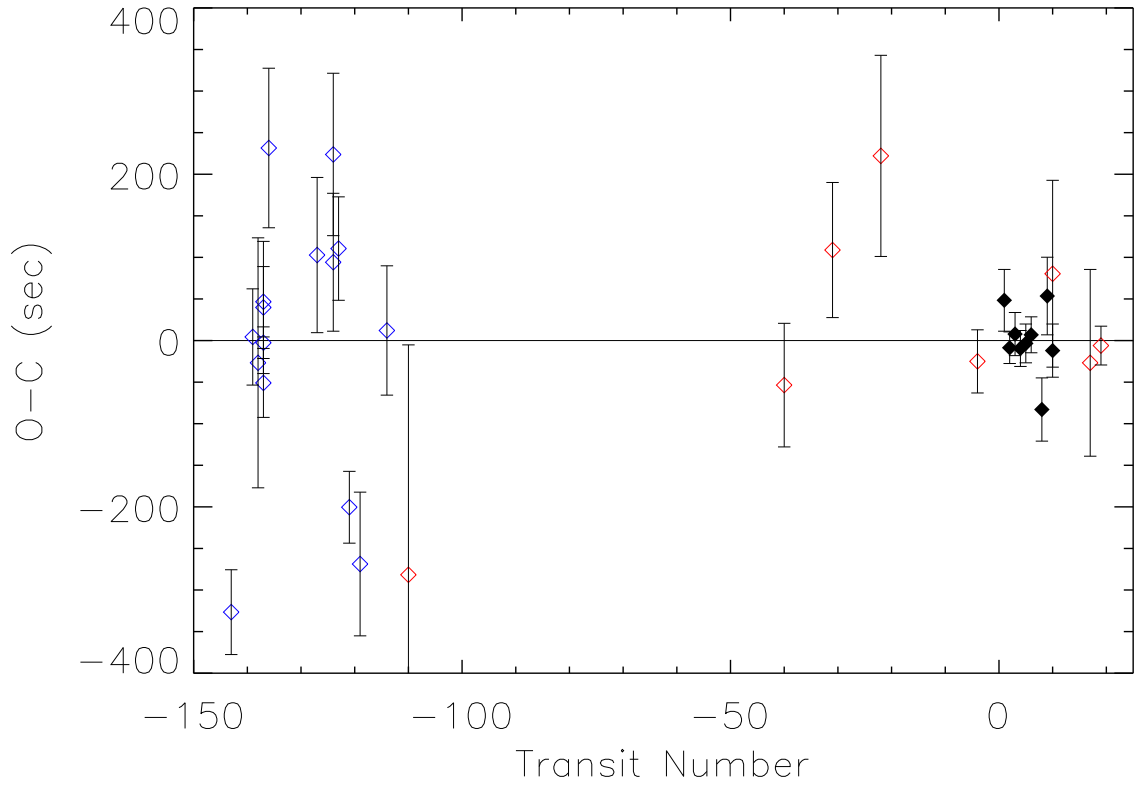}
\caption{Deviation from predicted time of transit vs. transit number for
        all published well sampled transits of HD 189733b.  Data points 
	from {\it MOST}
	are shown with filled symbols, those from \citet{win06b} with open red
	symbols, and those from
	\citet{bak06} with open blue symbols.  Note that different
	methods have been used to compute the error bars for each individual 
	data set.  See the original papers for details. 
        \label{O-C_all}}
\end{figure}

The resulting transit times are listed in Table~\ref{tbl2}.  We show these
measurements in an O-C (Observed time - Computed time) diagram in 
Figure~\ref{O-C_2006}, where the expected 
time of each transit (for constant orbital period) has been subtracted from 
its observed time to search for
transit timing deviations.  The expected times of transit are calculated
from the period and ephemeris that we present in Section~\ref{analysis}.
In Figure~\ref{O-C_all} we once again present an O-C diagram of the 
transit times for HD 189733b but now including, as a reference, all available 
timing data from the literature in addition to our own.  

For the first 6 transits observed by {\it MOST}, the transit
times we obtain are consistent with a constant orbital period at the 
1-$\sigma$ level.  The final 3 transits 
show increased scatter and uncertainty due to the fact that the light curve 
does not fully sample those transits, and portions of either ingress or 
egress are not observed.  Gaps
in the light curve during times when the planet is passing over the
limb of the star can have a large effect on the accuracy to which transit
times can be determined.  This is due to the fact that fits to partial 
transits are highly sensitive to the values used for the 
limb-darkening parameters and the orbital inclination.  If any of these 
deviate from their true values
by even a small amount, the effect on the transit time 
can be substantial (see \citet{mil07}).  We do not use the last three 
transits for our subsequent TTV analysis, as we expect the formal error bars 
in these cases to be underestimates.

\section{Transit Timing Variations in the HD 189733 System}\label{analysis}

The transit timing data on HD 189733 from {\it MOST} show no variations
on three scales: (a) no long-term change in $P_{\rm orb}$ in about
15 years at the 60 ms level; (b) no trend in transit timings during
the first two weeks of the {\it MOST} run; and (c) no individual transit 
timing deviations at the 45 s level, with a string of 5 consecutive transits
showing no timing deviations larger than 10 s. For illustration see 
Figures~\ref{O-C_2006} and~\ref{O-C_all}.

\subsection{Long-term Variations in the Orbital Period of HD 189733b}

We first address point (a) from the list above. Long-term variations in
the orbital period could be caused by several effects, e.g., orbital
decay of the planet HD 189733b \citep{sas03}, or precession of its orbit
\citep{mir02, hey06}.
By combining archival Hipparcos photometry from
1991 and 1993 with the discovery data of \citet{bou05}, \citet{heb06} 
have measured the orbital period of HD 189733b over a 15-year baseline to be
$2.218574^{+0.000006}_{-0.000010}$ d.  
\citet{win06b} independently determined the orbital 
period for HD 189733b to be 2.2185733 $\pm$ 0.0000019 days.  
Their measurement was based on a compilation of all available transit 
times for fully 
observed transits from their own work as well as from \citet{bak06}.

To further refine the value for the orbital period of HD 189733b and also
to reduce the size of its associated error bar, we present here a new
calculation of the period.  We follow the same method as 
\citet{win06b} and use all available timing data for full transits of
HD 189733b, adding our own data points from the first six transits observed
by {\it MOST} to the list.  We fit the resulting 18 transit times with a 
linear function of the form  
\begin{equation}
T_{c}(N) = T_{c}(0) + PN \ ,
\end{equation}
where $T_{c}(0)$ is the epoch and $N$ is the transit number.  
We choose $N=0$ as the fourth transit observed by 
{\it MOST}, since this is the transit time known to the highest precision.  
The orbital period and epoch obtained 
from the fit are 2.2185733 $\pm$ 0.0000014 d and JD 2453955.52478 $\pm$
0.00010, respectively.  Our new value for the period is entirely consistent 
with all previous determinations.  We have however, 
reduced the size
of the 1-$\sigma$ error bar to 120 ms, now making this the most accurate 
value available for the orbital period of HD 189733b.  


The star HD 189733 is estimated to have a mass of 0.82 $M_{\odot}$ with 
about 3\% uncertainty
\citep{bou05, bak06}. With our stellar model for this mass and a planet 
mass of 1.15 $M_{Jup}$, we can compute orbital decay rates.  For the
fastest possible orbital decay (high dissipation linear model 
\citep{zah77,zah89}), we get a rate 
of 0.6 msec per year. Currently available transit timing data for HD 189733b
are not yet sensitive to long-term changes in $P$ at this level.  However, 
with the accumulation of future transit timing observations of this sytem, 
further constraints can be placed on this estimate.

\subsection{Short-term Variations in the Orbital Period of HD 189733b Due to Additional Close-in Planets in the System}

The existing radial velocity curve for HD 189733 \citep{win06} excludes
additional periodic variations (other than the known giant planet) 
larger than 10 m/s in amplitude and up to a period of about 40 d. This
puts a limit on the mass of a possible long-period perturbing planet at
about $M_{\rm pert}\gtrsim 10^{-4}M_{\odot} \simeq 0.1 M_{Jup} \simeq 32
M_{\earth}$.  The TTV method is complimentary to the RV method in that it is 
generally more sensitive
to smaller close-in perturbing planets.  This is especially true in the case 
of the additional planet lying in a resonant orbit with the transiting planet, 
where the interactions between the two bodies are strongest.  We use only the 
first 6 {\it MOST} transits here (see $\S 3.2$), so the run can constrain only 
short-period perturbers in nearby resonances to HD 189733b, and periods 
shorter than about 6 days. 

To determine what types of planets are ruled out by the {\it MOST} 
data, both in mass and orbits, we solve the classical N-body problem,
\begin{equation}
\frac{d^2x_i}{dt^2} = - \Sigma_{j=1;j\neq i}{N} \frac{Gm_j (x_i -
x_j)}{|x_i-x_j|^3},
\end{equation}
where for 3 bodies, x describes the initial positions of the
particles.  A third body was inserted with an initially
circular orbit, with periods ranging from 1 to 9 days in increments of
0.01 days, with masses from 1 - 100 $M_\earth$ in 1 $M_\earth$ increments, and 
on a coplanar orbit to HD 189733b.  
The solution was advanced at 1.0 second intervals for 100 orbits of HD
189733b ($\sim 2\times 10^{7}$ s) using the LSODA routine from ODEPACK
\citep{rad93}.  O-C values were then calculated by a
linear interpolation to estimate the integration time when 
HD 189733b returns to the midpoint of crossing the disk of the star.  

\begin{figure}
\plotone{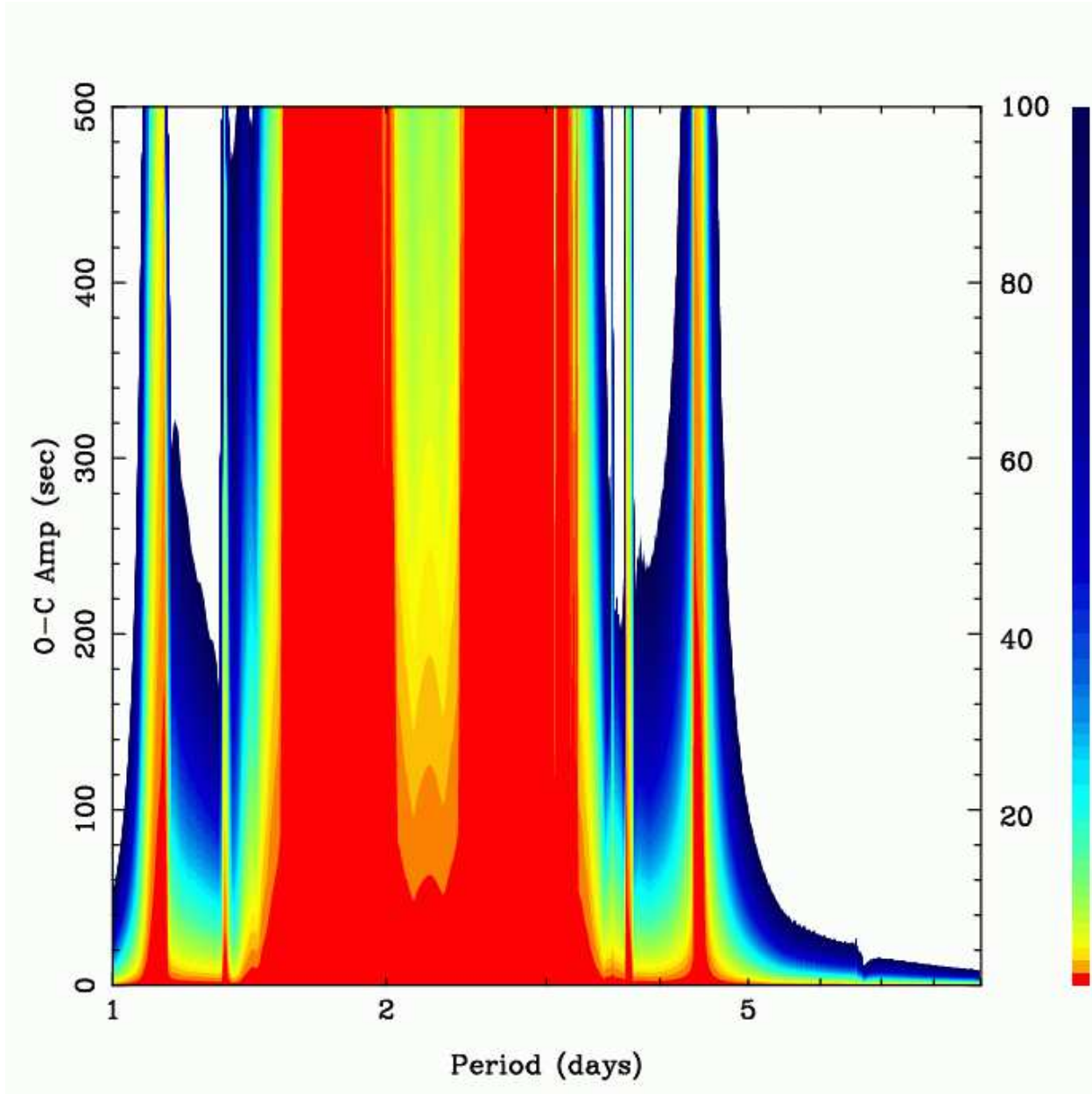}
\caption{N-body results for maximum transit timing deviation vs.\ orbital
        period of the perturbing planet.  The color scale as defined on the
        right side of the plot indicates the mass of the perturbing planet
        (in M$_{\earth}$).  The {\it MOST} data, covering 6 orbits of HD 
	189733b, show no transit timing variations above the level of 45 s.
	\label{JR_figure}}
\end{figure}

To determine the limits that transit timing data can place on additional 
planets in the HD 189733 system, we
compute a Fourier Transform of the O-C series generated by the N-body code 
for each value of period and mass of the 3rd body.  We then extract the 
largest amplitude as shown in Figure~\ref{JR_figure}.
However, we note that the largest amplitude of TTV can only be recovered from 
transit
timing data if the entire libration period of the two-planet system is fully
sampled.  Otherwise, there is a risk that the times of largest transit timing
deviations could be missed or passed over by the observations.  
Figure~\ref{JR_figure}
therefore shows what types of limits on companion planets could be achieved
if data were available for the full libration period of each hypothetical 
two-planet system.  Given that the {\it MOST} transit timing data only 
cover 6 orbits of HD 189733b, we use the following procedure to
determine what additional planets are ruled out from the system for cases 
where the libration period is longer than the two-week duration of the 
{\it MOST} observing run.   

\begin{figure}
\plotone{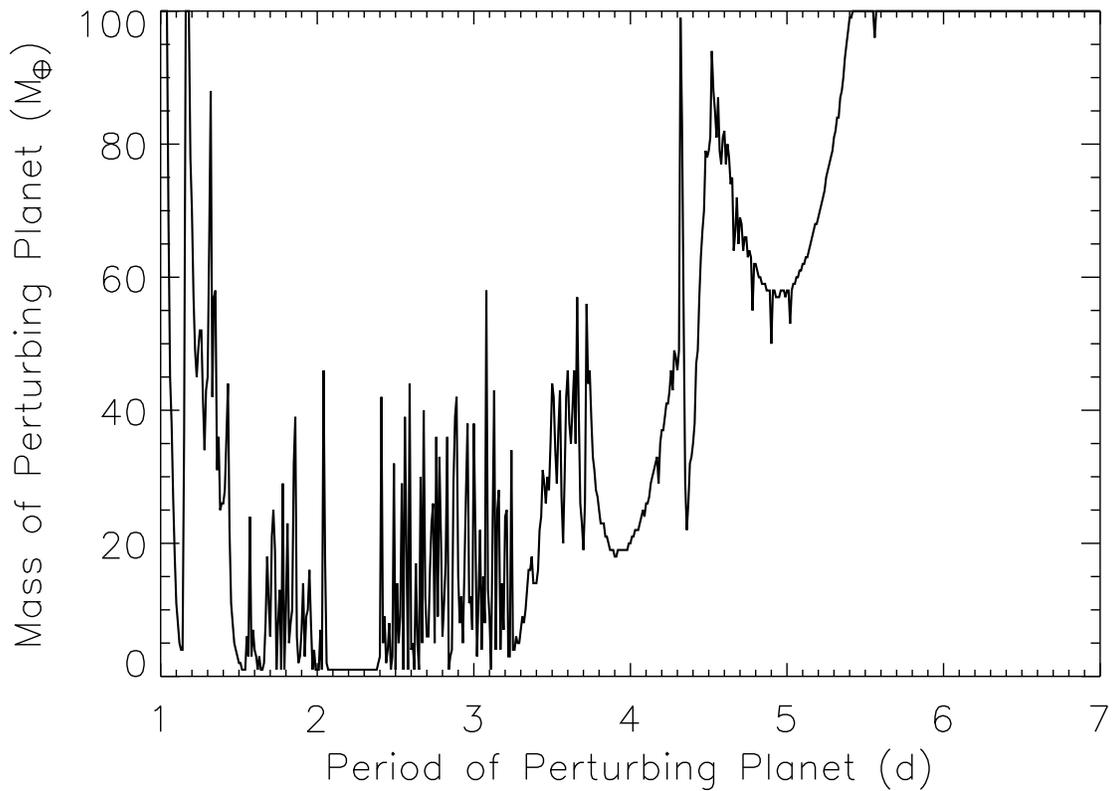}
\caption{Maximum mass allowed for a perturbing planet in the HD 189733 system, 
        which still remains consistent with the {\it MOST} transit times.  
	Planets occupying the region of parameter space above the curve are 
	ruled out by the available transit timing data with 95\% confidence.
        \label{EMR_figure}}
\end{figure}

We compare the N-body results against the {\it MOST}
transit timing data to determine the maximum mass that an additional planet 
in the system can possess, while still remaining consistent with the data. 
This is achieved by removing any linear trends from each of the O-C series 
generated by the N-body code,
and then optimizing the agreement with the observed transit timing residuals by
minimizing the chi-squared statistic.   
We then allow for the computed transit times to shift by an integer number of 
transits and determine the minimum value for chi-squared once the overall 
best fit is achieved.  For each configuration, the perturbing planet is ruled 
out if 
the minimum in chi-squared is higher than a set threshold, implying that the 
results from the N-body simulation are inconsistent with the 
{\it MOST} transit times.  Figure~\ref{EMR_figure} shows the results
of this process.  The maximum mass of perturbing planet that remains 
consistent with the transit timing data at a confidence level of better than 
5\% is 
plotted as a function of the orbital period of the perturbing body.  This
corresponds to a chi-squared of less than 11.07 for the best fit of 
the N-body simulation to the data, on 5 degrees of freedom.

According to Figure~\ref{EMR_figure}, the TTV limits exclude perturbers of 
greater than 20 $M_{\earth}$ for the exterior 2:1 mean motion resonance and 
greater than 8 $M_{\earth}$ in the 3:2 resonance.
In the interior 1:2 and 2:3 resonances, additional planets with masses larger 
than 4 $M_{\earth}$ and 1 $M_{\earth}$ are ruled out, respectively.  
A range of intermediate orbits near the transiting planet HD 189733b
are unstable for the lowest-mass perturbers, resulting in ejections or
collisions.  Between 2.0 and 2.4-day periods we can rule out sub-Earth
mass planets due to both TTV constraints and stability requirements.  
Additionally, we can place more stringent limits than the 32 $M_{\earth}$ value
from radial velocity measurements in a number of non-resonant orbits with 
periods of up to 5 days.

For a perturbing planet on an eccentric orbit, the TTV's induced
in the orbit of HD 189733b can be significantly larger than for the case of
a circular orbit.  Also, there is a reduced range of stable 
configurations for planets on eccentric orbits, especially for orbits interior
to that of HD 189733b.  We have performed a limited number of additional 
N-body simulations for pertubing planets in both  mutually inclined orbits 
relative to the transiting planet, and in initially eccentric orbits.  In 
almost all 
of these cases, we can place even stronger limits on the presence of additional
planets in the HD 189733 system.  In this sense, the limits that we have 
placed on perturbing planets from Figure~\ref{EMR_figure} are robust limits 
across the entire range of eccentricity parameter space, since we initially 
only considered planets with zero eccentricity.  Additional planets
residing in eccentric or inclined orbits would have even larger effects on
the transit times of HD 189733 than what we reported above.

\section{Considerations Due to Star Spots}

The amplitude of the intrinsic stellar variations during the {\it MOST} 
observing run suggests that HD 189733 may have had fairly large star spots of
high flux contrast during this time.  It is therefore quite possible that
the exoplanet may have transited one or more spots during the 21-day run.
This type of event
would alter the shape of the transit lightcurve in a predictible way, and we
can look for telltale signs of such an occurrence.  
From deviations in the
shape of the transit light curve, \citet{pon07} observed 
HD 189733b occulting starspots in photometry from the HST. The approximate 
change in signal was 0.7\%, and thus they inferred the presence of spots at
least 80,000 km across (1.1 R$_{Jup}$), given an inferred temperature of the 
spots that was 1,000 K cooler than the star itself.

Were the transiting planet to have crossed over a star spot during the 
{\it MOST} observing run, the depth of
the transit at this time would be shallower than otherwise predicted
due to the fact that the planet would be passing in front of a cooler, hence
optically dimmer, region 
of the photosphere.  Since the lessening in transit depth would occur 
only during the portion of the transit when the planet is actually in front
of the star spot and not necessarily during the entire transit, the overall
effect when fitting the \citet{man02} transit model to the light curve would
be twofold.  First, the shape of the transit would deviate from the 
form predicted by the transit model, resulting in a poorer goodness-of-fit
($\chi^2$) value.  Secondly, a shallower 
than normal overall depth would be needed to best describe a transit affected
by a star spot.  

Taking into account the second effect, we have examined the {\it MOST} data
for variations in depth between successive 
transits to determine if such a spot-crossing event took place during the 
{\it MOST} observing run.  For each of the six
fully observed transits, we now allow for the depth of the transit, 
$(R_{p}/R_{*})^{2}$, to be a free parameter in our fit.  While
the standard \citet{man02} template transit model cannot predict the correct 
transit shape in the case where the planet crosses a star spot, this analysis 
still allows for an estimate as to whether there are significant changes in 
the average transit depth throughout the {\it MOST} observing run.   To 
determine the size
of the error bars of these measurements, we employ the same bootstrap method 
described above for the transit timing measurements.  This
once again allows us to account for correlated noise in the data, which
can also affect the transit depth.  The resulting
values for transit depth are given in Table~\ref{tbl3}.  The mean transit 
depth is 2.660\% with a mean uncertainty of ${\pm}0.027\%$.  All six
measurements are consistent with a constant depth at the 2-$\sigma$ level.  

With the current precision in our transit depth measurements, we note that
we cannot detect changes in depth smaller than about 3\% at 
the 3-$\sigma$ level.  Essentially, this means that the noise in the
data could conceal the passage of HD 189733 over star spots smaller than 
approximately 0.2 R$_{J}$ in radius, under the assumption that the star 
spots are 1,000 K cooler than the rest of the photosphere 
(see discussion below).  It is entirely possible that this type of
small spot is present on the surface of HD 189733.  

\citet{cro07} have obtained the best fit of the unfiltered {\it MOST} 
light curve to a simple spot model with two large star spots, 
obtaining a rotation 
period of $11.73^{+0.07}_{-0.05}$ d and a stellar rotational inclination of 
$59^{+3}_{-8}$ degrees.  The spot model period is very close to 
the rotation period of $11.953^{+}_{-}0.09$ d determined by \citet{hen08} from 
groundbased photometry spanning from October, 2005 to July, 2007.  However,
the two values still differ by several sigma; a fact that may point to 
differential rotation for the set of star spots that was being observed. 
The \citet{cro07} model for HD 189733 assumes circular spots, solid-body 
surface rotation, and the possibility of evolution in spot size during the 
observations.  The star spots are assumed to be 1,000 K cooler than the 
rest of the photosphere, which is consistent with Sun spots and 
with HST observations of the HD 189733 system \citep{pon07}.  However, the 
star spot temperature cannot be determined 
uniquely from the {\it MOST} light curve, and there is a degeneracy between 
the sizes and temperatures of the spots.     

As a demonstration of the power of studying transits in systems with spotted 
stars, we combine the {\it MOST} transit observations with the star spot 
analysis of \citet{cro07} to predict whether HD 189733b should have 
transited in front of one of the modeled star spots during the observing 
run. The spot model predicts the position and size of the two large star 
spots on the disk of HD 189733 at any time during the {\it MOST} run.  

Using snapshots generated from the spot model discussed above, we can 
overlay the position of the transiting planet on an image of the stellar 
disk to see if HD 189733b is expected to have passed over
either of the two large star spots, as shown in Figure~\ref{spots}.  In this
figure, it is possible to project the position of the planet on the star 
by the combination of three independent pieces of
information.  First, the sky projection angle between the orbital axis of
the transiting planet and the stellar rotation axis of the star is known
to be $\lambda = -1.4^{\circ} \pm 1.1^{\circ}$ \citep{win06}.  This value was 
determined by measuring the Rossiter-McLaughlin effect, which is
the radial velocity distortion seen as a transiting planet occults 
the projected rotation profile of the star (see 
\citet{oht05, gim06, gau07}).  In addition, 
the stellar inclination angle ($I_{*}$) has been obtained from combining the 
measurement of $v \sin I_{*}$ (also from the Rossiter-McLaughlin effect)
\citep{win06} with the rotation period of star obtained by \citet{cro07}.  
Finally, the inclination angle of the planet's orbit is known
from fits to the transit light curve.  The only remaining
ambiguity lies in the fact that
it is not known whether the transit crosses along a chord below or above the
stellar equator.  In Figure~\ref{spots}, we depict both cases. 

From Figure~\ref{spots}, it can be seen that if HD 189733b were positioned 
above the
stellar equator as it transits, it would have crossed one of the large star
spots in the \citet{cro07} model during 4 out of the 6 fully-observed 
transits.  A complete occultation of a star spot 
comparably sized to the transiting planet would reduce 
the transit depth to 35\% of its original value.  
Such a large effect would be readily noticeable by eye, and it 
has also been ruled out by the transit depth analysis presented above for the
first 6 transits from the {\it MOST} light curve.  We can therefore conclude
that, if the \citet{cro07} model is correct, 
HD 189733b must transit below the stellar equator (as depicted in 
Figure~\ref{spots}).  In this case, the planet is not expected
to cross over either of the large star spots, with the possible exception of 
Transit 3.  In this single case, the combined uncertainties in the spot
model and the orbital inclination axis of the planet may allow for HD 189733b 
to transit across the southern hemisphere spot.  There is no evidence in
the light curve of Transit 3 to suggest a full occultation of this star spot.
However, a partial occultation cannot be ruled out at the precision of the
{\it MOST} data, making this transit the most likely one to be affected
by the presence of star spots as determined by \citet{cro07}.   

As an additional consideration, in the event that the transiting 
planet crosses in front of a spot during 
ingress or egress, the change in depth of the transit
from its expected value would be minimal.  However, since such an event
would change the shape of ingress or egress, the effect on the 
measured transit time would be considerable as this is where the timing signal
is most sensitive.  This is not expected to 
have occured in the HD 189733 system during the {\it MOST} observing run 
if the \citet{cro07} spot model is correct
(see Figure~\ref{spots}).  However, 
a transit across a star spot during ingress or egress could 
certainly take place during future observations.  To determine the expected 
change to the transit time if such an event were to occur, we placed a
simulated spot-crossing event into the data for one of the {\it MOST} 
transits.  We found that a transit across a dark star spot with a diameter of 
0.5 R$_J$ during ingress could alter the measured transit time by more than
a minute.  Such an event remains virtually undetected by eye in the {\it MOST}
data, but could be detected with spacebased photometry of sufficient time 
sampling and precision such as that of CoRoT, Kepler, or the HST.  

\section{Discussion and Conclusions}

We have measured accurate transit times for 9 transits of HD 189733b observed 
by {\it MOST} during 21 days in August, 2006.  For the 6 consecutive fully 
sampled transits
we find no transit timing deviations above the 45-s level, allowing
us to rule out certain configurations of additional perturbing planets in the 
system.  We find no evidence for perturbing planets in the 1:2, 2:3, and 3:2 
resonances down to the Super-Earth level (4, 1, and 8 $M_\earth$ 
respectively).  In the exterior 2:1 resonance we rule out perturbing planets
above the 20 $M_\earth$ level, and we place the most stringent limits 
available on a range of non-resonant orbits with periods ranging from 1 to 5
days. In addition, we find no evidence for long-term changes in the orbital 
period of HD 189733b between 1991 and 2006, indicating no significant 
observable orbital decay or precession.  

While HD 189733b is currently the only exoplanet known to transit a 
chromospherically active star, many more such systems 
may be discovered in the coming years,
especially with the launches of the CoRoT and Kepler space missions.  The 
occultation of a star spot changes the shape and depth of 
an observed transit, and passage of an exoplanet over a star spot during 
transit ingress or egress changes the measured center time of the transit.  
This is of particular concern for lower-precision photometry, where such 
deviations in the shape of the transit light
curve may not be readily detectible, but the effects on the transit timing
measurement can be significant.   

For the transit timing analysis of the {\it MOST} light curve for HD 189733, 
we have been able to account for the presence of star spots in several
ways.  (1) The nearly continuous 21-day coverage and high precision of the
data make it possible to subtract the intrinsic stellar variability 
before performing the transit timing analysis.
(2)  We have tested and constrained a two-spot model for
the longer-term variations in the {\it MOST} light curve \citep{cro07}, 
and combined it with spin-orbit alignment information from Rossiter-McLaughlin 
effect measurements  \citep{win06}.  This has allowed us to project
when and where the transiting planet would be expected to pass over a 
modeled star spot.  (3) Our error analysis for the transit times takes into 
account correlated 
noise in the {\it MOST} light curve, due either to possible residual 
instrumental effects, or to passages of HD 189733b over small spots 
on the star that are below the detection threshold for this data set.

Transiting planetary systems such as HD 189733 point to the intriguing 
possibility that the transits themselves could be useful probes of the 
nature of spots and activity on the host star.  High-quality photometry of 
future transits of this system could be used to 'map'  the surface brightness 
and geometry of spots in the part of the stellar disk traversed by the 
transiting planet.  

We urge any observers reporting transit times for planets transiting active 
stars to carefully consider possible star spot effects in any observed 
deviant transit times before assuming the presence of a perturbing body in 
the system. The type of analysis presented in this paper can be implemented 
in principle for any system to rule
out star spots as a cause of TTVs.  This method necessitates a light curve
taken with high photometric precision and complete time coverage over
at least two stellar rotation periods, to detect and model the low-level 
variability of the star due to the motion of star spots across its surface.  
Such observations
are best obtained from space or with a network of medium-aperture
groundbased telescopes such as the planned Las Cumbres Observatory Global 
Telescope \citep{bro06}.  Such data will help to make
headway in the search for low-mass planets in known transiting systems,
even in the case where the stars are chromospherically active.  

\acknowledgements

\begin{figure}
\includegraphics[scale=0.59]{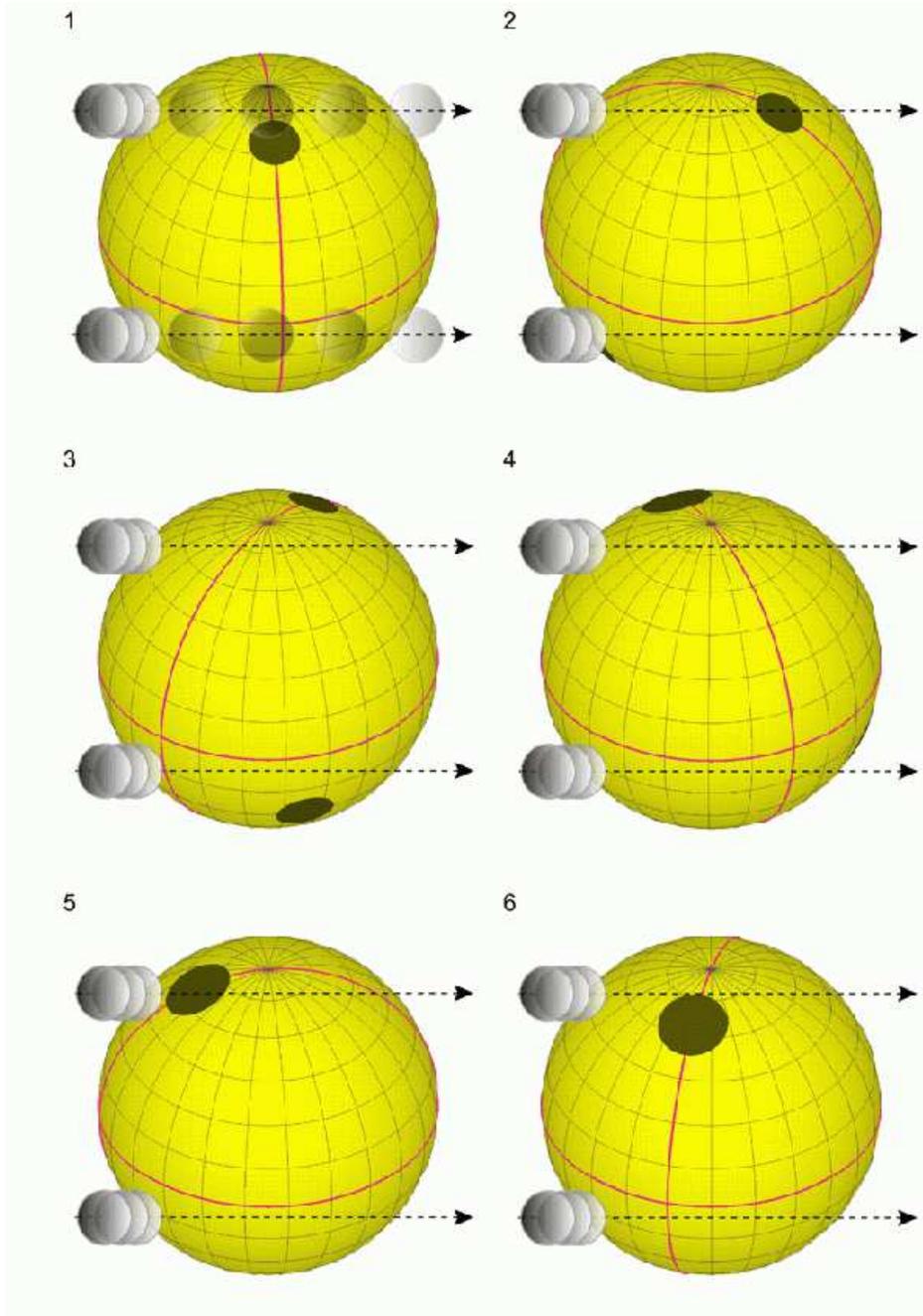}
\caption{Snapshots from the HD 189733 star spot model by \citet{cro07}, 
        predicting the position of two large star spots at the times of the
	6 complete transits observed by {\it MOST}.  The projected tracks of 
	the transiting planet are also shown, with two tracks for each transit 
	since only the orbital inclination, not the direction (north vs. 
	south) is known.  A northern hemisphere transit is ruled
	out if the Croll et al. model is correct, since effects of the 
	passage over the large star spot in this region
	would be readily visible in the transit light curve.
        \label{spots}}
\end{figure}

We would like to thank Paul A. Kempton for his help in the graphical design of 
Figure~\ref{spots}. JMM, DBG, AFJM, and
SMR acknowledge funding from the Natural Sciences \&
Engineering Research Council (NSERC) Canada.  AFJM is also supported by 
FQRNT (Quebec).  RK is supported by the Canadian
Space Agency.  WWW is supported by the Austrian Research Promotion Agency 
(FFG) and the Austrian Science Fund (FWF P17580).

\bibliography{ms}

\begin{deluxetable}{lc}
\tablecaption{Orbital and Physical Parameters for HD 189733 \label{tbl1}}
\tablewidth{0pt}
\tablehead{
\colhead{Parameter} & \colhead{Value}}
\startdata
M$_{*}$ [M$_{\sun}$] &  0.82 $\pm$ 0.03\tablenotemark{a}  \\
R$_{*}$ [R$_{\sun}$] & 0.749 $\pm$ 0.009 \\
R$_{pl}$ [R$_{Jup}$] & 1.192 $\pm$ 0.019 \\
i [$^o$] & 85.70 $\pm$ 0.11 \\
P [days] & 2.2185733 $\pm$ 0.0000019\tablenotemark{b} \\
c$_{1}$ & 0.825232 \\
c$_{2}$ & -0.977556 \\
c$_{3}$ & 1.686737 \\
c$_{4}$ & -0.657020 \\

\enddata

\tablenotetext{a}{From \citet{bou05}}
\tablenotetext{b}{From \citet{win06b}}
\tablecomments{c$_{1}$-c$_{4}$ are the non-linear limb-darkening coefficients 
		calculated for the {\it MOST} bandpass.}

\end{deluxetable}

\begin{deluxetable}{rrrr}
\tablecaption{Best-Fit Transit Times \label{tbl2}}
\tablewidth{0pt}
\tablehead{
\colhead{Transit \#} & \colhead{T$_C$ (HJD)} & \colhead{$\sigma$ (HJD)} & \colhead{Reduced $\chi^2$}}
\startdata
1 &  2453948.86962 & $\pm$0.00043 & 0.82 \\
2 &  2453951.08753 & $\pm$0.00022 & 0.83 \\
3 &  2453953.30630 & $\pm$0.00030 & 0.88 \\
4 &  2453955.52467 & $\pm$0.00025 & 0.61 \\
5 &  2453957.74331 & $\pm$0.00027 & 0.76 \\
6 &  2453959.96201 & $\pm$0.00025 & 0.65 \\
8 &  2453964.39811 & $\pm$0.00044 & 1.12 \\
9 &  2453966.61827 & $\pm$0.00054 & 1.00 \\
10 &  2453968.83608 & $\pm$0.00037 & 0.97 \\
\enddata

\tablecomments{Transit 7 has been omitted due to a large gap in the data at 
         this time.}

\end{deluxetable}

\begin{deluxetable}{rrrr}
\tablecaption{Transit Depth Measurements\label{tbl3}}
\tablewidth{0pt}
\tablehead{
\colhead{Transit \#} & \colhead{Depth (\%)}}
\startdata
1 &  $2.661 \pm 0.048$ \\
2 &  $2.605 \pm 0.015$ \\
3 &  $2.631 \pm 0.019$ \\
4 &  $2.670 \pm 0.028$ \\
5 &  $2.701 \pm 0.023$ \\
6 &  $2.694 \pm 0.021$ \\
\enddata

\end{deluxetable}


\end{document}